\documentclass{aa}  

\usepackage{graphicx}
\usepackage{txfonts}
\newcommand{\adamemail}{adam.hardy@postgrado.uv.cl}

\newcommand{\Msun}{M$_{\odot}$}

\newcommand{\Mjup}{M$_{\mathrm{J}}$}

\begin{document}

\title{Probing the final stages of protoplanetary disk evolution with ALMA}

\author{A. Hardy\inst{\ref{valpo},\ref{mad}}  
\and C. Caceres \inst{\ref{valpo},\ref{mad}} \and M.R. Schreiber \inst{\ref{valpo},\ref{mad}} \and L. Cieza \inst{\ref{udp},\ref{mad}} \and R.D. Alexander \inst{\ref{ul}} \and H. Canovas \inst{\ref{valpo},\ref{mad}} \and J.P. Williams \inst{\ref{uh}} \and Z. Wahhaj \inst{\ref{eso}} \and F. Menard \inst{\ref{france}}}

\institute{
Departamento de F\'{i}sica y Astronom\'{i}a, Universidad Valpara\'{i}so, Avenida Gran Breta\~{n}a 1111, Valpara\'{i}so, Chile.\label{valpo} 
\and
Nucleo de Astronomia, Universidad Diego Portales, Av. Ej\'{e}rcito 441, Santiago, Chile.\label{udp}
\and
Department of Physics \& Astronomy, University of Leicester, University Road, Leicester, LE1 7RH, UK.\label{ul}
\and
Institute for Astronomy, University of Hawaii, 2680 Woodlawn Drive, Honololo, HI, USA.\label{uh}
\and
European Southern Observatory, Casilla 19001, Vitacura, Santiago, Chile.\label{eso}
\and
UMI-FCA, CNRS/INSU, France (UMI 3386), and Dept. de Astronom\'{\i}a, Universidad de Chile, Santiago, Chile. \label{france}
\and
Departamento de Astronom\'{i}a, Universidad de Chile, Camino del Observatorio 1515, Las Condes, Santiago,  Chile. \label{calan}
\and
ICM nucleus on protoplanetary disks, Universidad de Valpara\'{\i}so, Av. Gran Breta\~na 1111, 2360102 Valpara\'{\i}so, Chile.\label{mad}
\begin{center}
\email{\adamemail}
\end{center}}
\date{Received <date> /
Accepted <date>}

\abstract
{The evolution of a circumstellar disk from its gas-rich protoplanetary stage to its gas-poor debris 
stage is not understood well. It is apparent that disk clearing progresses from the inside-out on a short time scale and models of photoevaporation are frequently used to explain this. However, the photoevaporation rates predicted by recent models differ by up to two orders of magnitude, resulting in uncertain time scales for the final stages of disk clearing.}
{{Photoevaporation theories predict that the final stages of disk-clearing progress in objects that have ceased accretion but still posses considerable material at radii far from the star. Weak-line T Tauri stars (WTTS) with infrared emission in excess of what is expected from the stellar photosphere are likely in this configuration. 
We  aim to provide observational constraints on theories of disk-clearing by measuring the dust masses and CO content of a sample of young (1.8-26.3 Myr) WTTS.}}
{We used ALMA Band 6 to obtain continuum and $^{12}$CO(2-1) line fluxes for a sample of 24 WTTS stars with known infrared excess. For these WTTS, we inferred the dust mass from the continuum observations and derived disk luminosities and ages to allow comparison with previously detected WTTS.}
{We detect continuum emission in only four of 24 {WTTS}, and no $^{12}$CO(2-1) emission in any of them. For those {WTTS} where no continuum was detected, their ages and derived upper limits suggest they are debris disks, which makes them some of the youngest debris disks known. Of those where continuum was detected, three are possible photoevaporating disks, although the lack of CO detection suggests a
severely reduced gas-to-dust ratio.}
{The low fraction of continuum detections implies that, 
once accretion onto the star stops, the clearing of the majority of dust progresses very rapidly. Most WTTS with infrared excess are likely not in transition but are instead young debris disks, whose dust is either primordial and has survived disk-clearing, or is of second-generation origin. In the latter case, the presence of giant planets within these WTTS might be the cause.}

\keywords{Protoplanetary disks -- Planets and satellites: formation --  Planet-disk interactions -- Radio continuum: planetary systems -- Radio lines: planetary systems -- Infrared: planetary systems}

\maketitle

\section{Introduction}

Circumstellar disks typically fall into two categories -- the massive gas-rich protoplanetary disks and the gas-poor debris disks. Protoplanetary disks are a natural consequence of the star formation process, with the majority of their evolution being dominated by viscous accretion onto the star{ \citep[e.g.,][]{Lynden-Bell1974}}. They can easily be  identified by their near and mid-infrared (mid-IR) excesses, caused by their vast optically thick dust disks \citep[e.g.][]{Strom1989}, {although the majority of their mass in fact resides in gas with typical gas-to-dust mass ratios of $\sim$100 \citep[e.g.][]{Pietu2005,Panic2008}}.  The debris disks, meanwhile, generally contain little or no detectable gas, with dust often confined to a narrow ring \citep[see][for a review]{Wyatt2008}. It has long been suggested that debris disks could be a later stage of evolution from the protoplanetary disks, but the nature of the main physical processes that
drive this evolution is ill-understood and remains one of the biggest questions in the
field. To understand this phenomenon, the nature of a class of disks
known as transition disks, needs to be investigated. 

While a unique and universally accepted definition of what constitutes a transition disk (TD) does not exist, the most general definition is that of a T-Tauri disk with reduced excess emission at near to far IR wavelengths relative to typical T-Tauri disks. Most TDs show significantly reduced fluxes at short wavelengths ($\lesssim$ 10 $\mu$m), while still showing average emission at longer wavelengths. This observation is indicative of a dust cavity in the innermost region of the disk, and indeed sub-mm images of TDs have proved this to be the case \citep{Pietu2006,Hughes2007,Brown2009,Canovas2015}. This kind of  geometry requires a mechanism that can clear the dust from the inside out, which is not consistent with the previously dominant mechanism of viscous accretion {\citep{Hartmann1998,Armitage1999}}. In addition, the lifetime of this transition stage must be short, i.e. $\leqslant$1 Myr because the detection rate of TDs is comparatively low, which again argues against the slow process of viscous accretion \citep{Wolk1996,Duvert2000, Andrews2005,Andrews2007}. Many mechanisms have been proposed to explain the phenomena of TDs, including grain growth, giant planet formation, binarity and photoevaporation. Although all these mechanisms probably contribute to disk evolution, photoevaporation is the most plausible mechanism
 to explain the final rapid removal of material at large radii.

Initial theories of photoevaporation \citep[e.g.][]{Hollenbach1994,Clarke2001} described the photoionisation of hydrogen in the disk surface by extreme ultraviolet (EUV) photons. This photoionisation forms a pressure gradient that is able to drive mass loss in a photoevaporative wind beyond a critical radius. Evidence to support this theory has been found {by comparing  models to} observations of both [NeII] \citep{Alexander2008, Pascucci2009, Ercolano2010, Pascucci2011, Sacco2012} and [OI] lines \citep{Font2004,Gorti2011,Rigliaco2013}.  
Although this process is likely to  occur throughout the disk's lifetime, it only becomes significant when the accretion rate becomes comparable to the photoevaporation rate. When this happens, photoevaporation can open a gap, forming an inner and an outer disk. The inner disk, now cut-off from re-supply, drains on a viscous time scale, thus creating the observed transition disk geometry \citep[see][for a review]{Alexander2014}. 
In the case of a transition disk however, the inner opacity hole allows radiation from the star to then reach the rim of the outer disk unimpeded, allowing it to complete the clearing of the disk on a short timescale of approximately $10^{5}$ years \citep{Alexander2006a, Alexander2007}. 

Although a very promising theory, EUV photoevaporation suffers from an uncertainty in the EUV flux incident on the disk. Stellar EUV flux is difficult to measure as interstellar absorption prohibits direct observation
of the ionising photons. Furthermore, these ionising photons can be blocked by optically thick accretion columns, jets or winds, resulting in final estimates for the flux that can reach the disk varying by orders of magnitude \citep{Herczeg2007, Pascucci2012}. Some recent advances in this area have had some success using free-free emission from the disk to place limits on the EUV flux, but the results suggest EUV wavelengths are, in fact, not sufficient to explain the [NeII] emission seen in some systems \citep{Pascucci2014}. 

{More recently, the effect of both X-ray \citep{Owen2010,Owen2011,Owen2012} and far ultraviolet (FUV) \citep[][]{Gorti2008,Gorti2009a,Gorti2009b} wavelengths have been included into photoevaporation models and a lot of debate surrounds the question of which wavelength is most dominant}. X-ray and FUV both predict the same evolutionary behaviour as EUV, but generally predict mass loss rates which are orders of magnitude higher than the original EUV models, with $\dot{M}_{PE} \sim 10^{-10}$ \Msun\, yr$^{-1}$ for EUV models, and $\dot{M}_{PE}\sim 10^{-8}$ \Msun\,yr$^{-1}$ for FUV/X-ray \citep{Gorti2009b}. Although stellar FUV and X-ray emissions are easier to measure than EUV, these models also suffer from uncertainties regarding disk chemistry and dust properties. FUV models, for example, describe heating of the gas being dominated by polycyclic aromatic hydrocarbons (PAHs), but the abundance and depletion of PAHs are difficult to determine \citep{Geers2009}.  

To answer some of the questions surrounding photoevaporation models, a study of transition disks in their last stages of gas clearing is required. A natural sample in which to find such disks is in weak-lined T-Tauri stars (WTTS). Unlike the classical T-Tauri stars (CTTS), WTTS have a narrow H$\alpha$ width, which is a strong indicator that the star is no longer accreting (or is accreting only at a relatively low level), and therefore lacks gas at radii close to the star \citep[see also][]{Pascucci2006, Ingleby2009}. The majority of WTTS also have no IR excess, suggesting they have already cleared all their circumstellar gas and dust and are not in transition. Nevertheless, a relatively small percentage of WTTS ($\sim20\%$) do display an IR excess, and this excess suggests a significant amount of dust (Cieza et al. 2007; Wahhaj et al. 2010). Even so,  a measurement of their dust mass has not yet been achieved and their nature as transition disks has not been confirmed. 

For example, in a dedicated survey of transition disks selected from Spitzer \citep{Cieza2010, Cieza2012,Romero2012} only one such WTTS 
system was detected at sub-mm wavelengths (FW\,Tau), but observations of this particular system with ALMA recently revealed that the sub-mm emission originates from around an accreting third object. FW\,Tau should therefore not be considered as a disk-possessing WTTS
\citep{Kraus2015,Caceres2015}.
 Furthermore, no previous studies have been able to measure the gas content of these disks, and  it is still to be determined whether WTTS with an IR excess (referred to as IR-WTTS in the remainder of this paper) are still in transition and photoevaporating their gas, or
whether they are more akin to young debris disks. If they should come under the former class, these systems will indeed be in the final stages of  gas clearing, and observational constraints on their mass and gas-to-dust mass ratio will be invaluable to theories of photoevaporation. If they should come under the latter class, then these disks will be among some of the youngest debris disks known, and their ages can provide a constraint on the initial conditions of the debris-disk phenomenon. Furthermore, if this result were to hold for a large sample of IR-WTTS, then this would suggest that by the time accretion ceases, the gas has either already been heavily depleted or the remaining gas has been photoevaporated rapidly. In this sense, such a result would have a strong influence on both photoevaporation theory and disk evolution in general.   

Here, we present one of the first detailed studies of WTTS with IR\ excesses using ALMA. We use Band 6, observing both the continuum and the $^{12}$CO(2-1) transition with the aim of identifying the evolutionary state of these WTTS disks and imposing limits on photoevaporation theory. 

\renewcommand{\arraystretch}{1.3}
\begin{table*}
\caption{Sample parameters}
\label{sampleparams}
\centering
\begin{tabular}{c c c c c c c c c c}
\hline\hline 
No. & 2MASS ID & Cloud & Distance & Spectral type &H$_\alpha$ width&  Ref. & Binary sep. & Binary ref. \\
 &  &  &  (parsecs) &&(km\,s$^{-1}$)&&(arsecs)&  \\
\hline 

1 & 04182147+1658470 & Taurus & 135$\pm$20 & K5 & Absorp. & 1,2,3  &... &...\\

2 & 04192625+2826142 & Taurus & 135$\pm$20 & K7 &180& 4,5 &{10.5}&1\\
 
3 & 04242321+2650084 & Taurus & 135$\pm$20& M2 &200& 5 &...&... \\

4 & 04314503+2859081 & Taurus & 135$\pm$20& F5   &Absorp.& 5  &...&...\\

5 & 04325323+1735337 & Taurus & 135$\pm$20 & M2 &138& 4,6 &...&...\\

6 & 04330422+2921499 & Taurus & 135$\pm$20 & B9 &Absorp.&  5 &...&...\\

7 & 04364912+2412588 & Taurus & 135$\pm$20 & F2 &Absorp.&  2,5 &...&...\\

8 & 04403979+2519061 & Taurus & 135$\pm$20 & M5   &130& 3,5 &...&...\\

9 & 04420548+2522562 & Taurus & 135$\pm$20 & K7 &94 & 3,6 &0.3&11\\

10& 08413703-7903304 & $\eta$Chamaleonis & 97$\pm$3 & M3 &Absorp.& 4,7 &...&...\\

11& 08422372-7904030 & $\eta$Chamaleonis & 97$\pm$3 & M3 &Absorp.& 4,7 &...&...\\

12& 11073519-7734493 & Chamaleon I & 160$\pm$15 & M4 &89& 1,8 &...&...\\

13& 11124268-7722230 & Chamaleon I & 160$\pm$15 & G8 &Absorp.& 1,8,9 &0.247&12\\

14& 16002612-4153553 & Lupus IV & 150$\pm$20 &   M5.25 &162& 10 &2.8&13\\ 

15& 16010896-3320141 & Lupus I & 150$\pm$20 &   G8 &Absorp.& 6 &...&...\\

16& 16031181-3239202 & Lupus I & 150$\pm$20 & K7 &132& 4,6 &...&...\\

17& 16085553-3902339 & Lup III  &  200$\pm$20 &  M6  &189&    10 &2.8&13\\ 

18& 16124119-1924182 & Ophiuchus & 119$\pm$6 & K8 &131&   6 &1.0&14\\

19& 16220961-1953005 & Ophiuchus & 119$\pm$6 &  M3.7  &132&  10 &1.8,3&10\\ 

20& 16223757-2345508 & Ophiuchus & 119$\pm$6 & M2.5 &128&  4,6 &...&...\\

21& 16251469-2456069 & Ophiuchus & 119$\pm$6 & M0   &206&  4,6 &...&...\\

22& 16275209-2440503 &  Ophiuchus & 119$\pm$6 & K7 & 151 & 4 & 0.480&15\\

23& 19002906-3656036 & Corona Australis & 129$\pm$11 &  M4 &93&  10 &0.132&16\\ 

24& 19012901-3701484 &Corona Australis & 129$\pm$11 &   M3.75 &83 & 10 &0.5&10\\

\hline
\end{tabular}

\tablefoot{The target's photometry, spectral types and H$_{\alpha}$ widths can be found in the following papers:- 1) \cite{Nguyen2012}; 2) \cite{Howard2013}; 3) \cite{Luhman2010}; 4) \cite{Cieza2013a}; 5) \cite{Cieza2012}; 6) \cite{Wahhaj2010};  7)
 \cite{Sicilia-Aguilar2009}; 8) \cite{Luhman2008}; 9) \cite{Matra2012}; 10) \cite{Romero2012}. The binary systems were identified in the following papers:- 11)  \cite{Leinert1993}; 12) \cite{Lafreniere2008}; 13) \cite{Merin2008}; 14) \cite{Prato2007}; 15) \cite{Ratzka2005}; 16) \cite{Koehler2008}.}
\end{table*}

\section{Observational procedure}

{Characterisation of the disks around IR-WTTS can be achieved by measuring  their dust and gas mass, stellar age, fractional disk luminosity and multi-wavelength photometry. It is these values that we therefore attempt to determine.}  

\subsection{Target selection}

Our sample consists of 24 pre-main-sequence stars (based on weak H$\alpha$ emission and Li I
absorption) in nearby ($\lesssim$ 200 pc) molecular clouds (see Table \ref{sampleparams}). The objects have been classed as a WTTS based on the velocity
  widths of their H$\alpha$ lines. The width of the H$\alpha$ line provides a reliable, distant-independent indication of a star's accretion, with accretion producing broad, asymmetrical H$\alpha$ emission. Non-accreting objects will still produce H$\alpha$ emission of chromospheric origin, but this tends to be comparatively narrow and symmetric. The empirical dividing line between accreting and non-accreting has been a matter of some debate, with some claiming that accreting systems have a 10$\%$ peak width of $>$270 km\,s$^{-1}$ \citep{White2003}, and others claiming that that the dividing line is $>$200 km\,s$^{-1}$, which varies with spectral type \citep{Martin1998}. The systems in this study were selected because they all either have  10$\%$ peak widths less than 200 km\,s$^{-1}$, or instead display H$\alpha$ absorption lines, allowing them to be identified as very likely non-accretors. 
  They all lack considerable IR excess at $\sim$10 $\mu$m or
   shorter wavelengths but show weak yet robust (>5-10 $\sigma$) excesses in the
    mid and/or far-IR from Spitzer, WISE, and/or Herschel. {These targets have been previously labelled
as either photoevaporating transition disks, or debris disks candidates, because of their lack of accretion and because of their estimated fractional disk luminosities having values less than the typical value for protoplanetary disks of $\sim 10^{-1}$ \citep{Wahhaj2010,Cieza2010,Cieza2012,Romero2012}. Those with fractional disk luminosities in the region of $10^{-1} \geq L_{D}/L_{\ast} \geq 10^{-2}$ were classed as photoevaporating transition disks, while those with a lower value were classed as debris disk candidates}. Binaries with wide projected separations were included, but no spectroscopic or confirmed close binaries were, because circumbinary disks are believed to undergo a different evolution and therefore need to be studied separately \citep{Kraus2012}. System 23 is a possible close binary, but only single epoch data is available for this system and its binary nature has yet to be confirmed via proper motion.

\subsection{ALMA observations}

\begin{table*}
\caption{Observation log}
\label{obslog}
\centering
\begin{tabular}{c c c c c c c c c c c}
\hline\hline 
Date & Cloud & Antennas & ToT & Antennas & BC & GC & PC & PWV & Min/Max\\
&&& (min) &flagged & (QSO) && (QSO) && baseline (m)\\
\hline 

12/06/17 & Lup/Op/CrA & 20   & 4.54 & ... & J1733-1304 & Neptune &   J1924-2914 &1.41 & 21.2/402\\

13/11/01 & Lup/Op & 29   & 2.07 & DV19,08,06 &  J1924-2914 & Neptune &  J1625-2527 &1.60&17.3/992 \\

13/12/04 & Taurus & 27   &  3.60 & DV08 & J0423-0120 & J0510+180 &  J0510+180 &5.11&15.8/463\\

13/12/18 & Cha & 23   &  4.16 & DV19,08 &   J1107-4449 & Ceres &  J0635-7516&2.31&15.1/992 \\

\hline

\end{tabular}
\tablefoot{{ToT - Time on target, BC - Bandpass calibrator, GC - Gain calibrator, PC - Phase calibrator, PWV - Precipitable water vapour}}

\end{table*}

Observations of the above systems were performed in Band 6, with Systems 14, 17, 19, 23 and 24 being observed in Cycle 0 (2012), and the remaining in Cycle 1 (2013). Cycle 1 observations were split into three, based on their host cloud, with the Lupus and Ophiucus systems in one group,  Taurus in another, and Chamaleonis in the final group. These groups were observed at different times and with slightly different configurations. Table \ref{obslog} gives a summary of these observational set-ups.

We chose to observe the $^{12}$CO(2-1) line since it is highly sensitive to the presence of circumstellar gas out to large radii, where the bulk of the gas should be located. We obtained one epoch of observation for all systems, with the correlator configured to obtain one baseband centred on 230.52 GHz which was aimed at detecting the $^{12}$CO(2–1) spectral line, and three continuum basebands centred at 228.52, 214.52, and 212.52 GHz. However, a fault with the local oscillator during the Cycle 0 observations meant that only the 230.52 and 228.52 GHz basebands could be observed. The total bandwidth for the observations was 3.75 GHz for the Cycle 0 and 7.5 GHz for Cycle 1 observations, with a unique spectral resolution of 976.56 kHz in 3840 channels for each 1.875 GHz baseband. In all cases, the requested rms was set at 0.16 mJy for the continuum, and 30 mJy for each individual velocity channel (and thus for the $^{12}$CO(2-1) line). Standard calibration steps were applied to the data, and  
we obtained the final images by deconvolving the set of visibilities with the clean task implemented in CASA \citep{McMullin2007}, using natural weighting. A point-source, fitted to the
measured visibilities, was used for those systems with a continuum detection to gain accurate estimation of this flux value.

\section{Results}

\begin{table}
\caption{Calculated parameters}
\label{results}
\centering
\setlength{\tabcolsep}{0.13cm}
\begin{tabular}{c c c c c c}
\hline\hline 

No. &  1.3mm Flux  & Dust mass & {Stellar mass} &  Age & $L_{D}/L_{\ast}$\\
   & (mJy) & (M$_{\oplus}$)& (M$_{\odot}$) & (Myr)  & ($\times10^{-3}$)\\
\hline 

1  &   $<$0.436 & $<$ 0.30 & ... & ... & $\leq$3.6 \\

2 &   0.533 & 0.28$\pm$0.09 & 0.67 & $3.5_{-1.1}^{+2.7}$&2.3 \\

3 & $<$0.426 &$<$ 0.30 & 0.35 &$6.7_{-1.6}^{+2.7}$ & $\leq$1.5 \\

4 & $<$0.435 &$<$0.30 &... &... &  $\leq$7.1 \\ 

5 &  $<$0.438 & $<$0.30 & 0.36 &$3.1_{0.6}^{+1}$& $\leq$3.5 \\

6 &   $<$0.431 &$<$0.30 & ... &...& $\leq$2.9\\

7 &  $<$0.435  &$<$0.30 & ... &...& 6.5\\

8 &  $<$0.431 &$<$0.30 & 0.17 &$4.1_{-0.7}^{+1.4}$ & $\leq$11\\

9 &  $<$0.423 &$<$0.29 & 0.65 &$1.8_{-0.5}^{+1.2}$ & $\leq$4.6\\

10  &  $<$0.476 & $<$0.14 & 0.27 &$9.0_{-0.6}^{+0.7}$ & $\leq$1.4\\

11 &  $<$0.472& $<$0.14 & 0.31 &$3.4_{-0.4}^{+0.5}$& $\leq$0.8\\

12 &  $<$0.541 & $<$0.48 & 0.195 &$15.0_{-3}^{+6}$ &  $\leq$2.3\\

13 &  $<$0.494 & $<$0.44 & 1.4 &$12.4_{-1.6}^{+2}$ & $\leq$2.3$^{\ast}$\\

14 &  0.696 &0.45$\pm$0.13  & 0.17 &$6.9_{-1.7}^{+2.8}$ & 80\\

15 &  $<$0.442 & $<$0.37 &  1.1 &$26.3_{-4.9}^{+7.4}$& $\leq$0.26\\

16&  $<$0.453 & $<$0.38 &  0.70 &$7.7_{-1.6}^{+2.3}$ &  $\leq$2.8\\

17& 1.813  & 2.09$\pm$0.44 &  0.10 &$2.6_{-0.5}^{+1.2}$ & 140\\

18&  $<$0.459 &  $<$0.21 &  0.64 &$6.6_{-1.5}^{+2.4}$ & $\leq$3.4\\

19&  $<$0.483&  $<$0.22 &  0.34 &$2.2_{-0.2}^{+0.3}$ & $\leq$5.95\\

20  &    $<$0.461 & $<$0.21 &  0.33 &$9.9_{-0.6}^{+0.8}$ &  $\leq$1.96\\

21 &   $<$0.453  &  $<$0.20 &  0.56 &$3.0_{-0.6}^{+0.4}$& $\leq$2.0\\

22 &   $<$0.442 & $<$0.20 &  0.77 &$4.2_{-0.8}^{+1.2}$& $\leq$1.5\\

23 & 0.569 &  0.27$\pm$0.05&  0.24 &$4.7_{-0.7}^{+1.0}$ & 28 \\

24&  $<$0.465 & $<$0.26 &  0.21 &$10.2_{-1.6}^{+2.3}$ & $\leq$5.7\\

\hline
\end{tabular}
\tablefoot{{Four systems have no calculated stellar mass or age as their luminosity does not coincide with any PMS tracks and are therefore probably background main-sequence stars. The final column of $L_{D}/L_{\ast}$ gives the value of fractional disk luminosity, calculated by integration of the system spectral energy distribution}. $^{\ast}$The value of $L_{D}/L_{\ast}\leq 10^{-3}$ quoted for system 13 assumes the far-IR emission is due to contamination.}
\end{table}

\subsection{Disk dust masses}

\begin{figure*}
    \resizebox{\hsize}{!}{\includegraphics{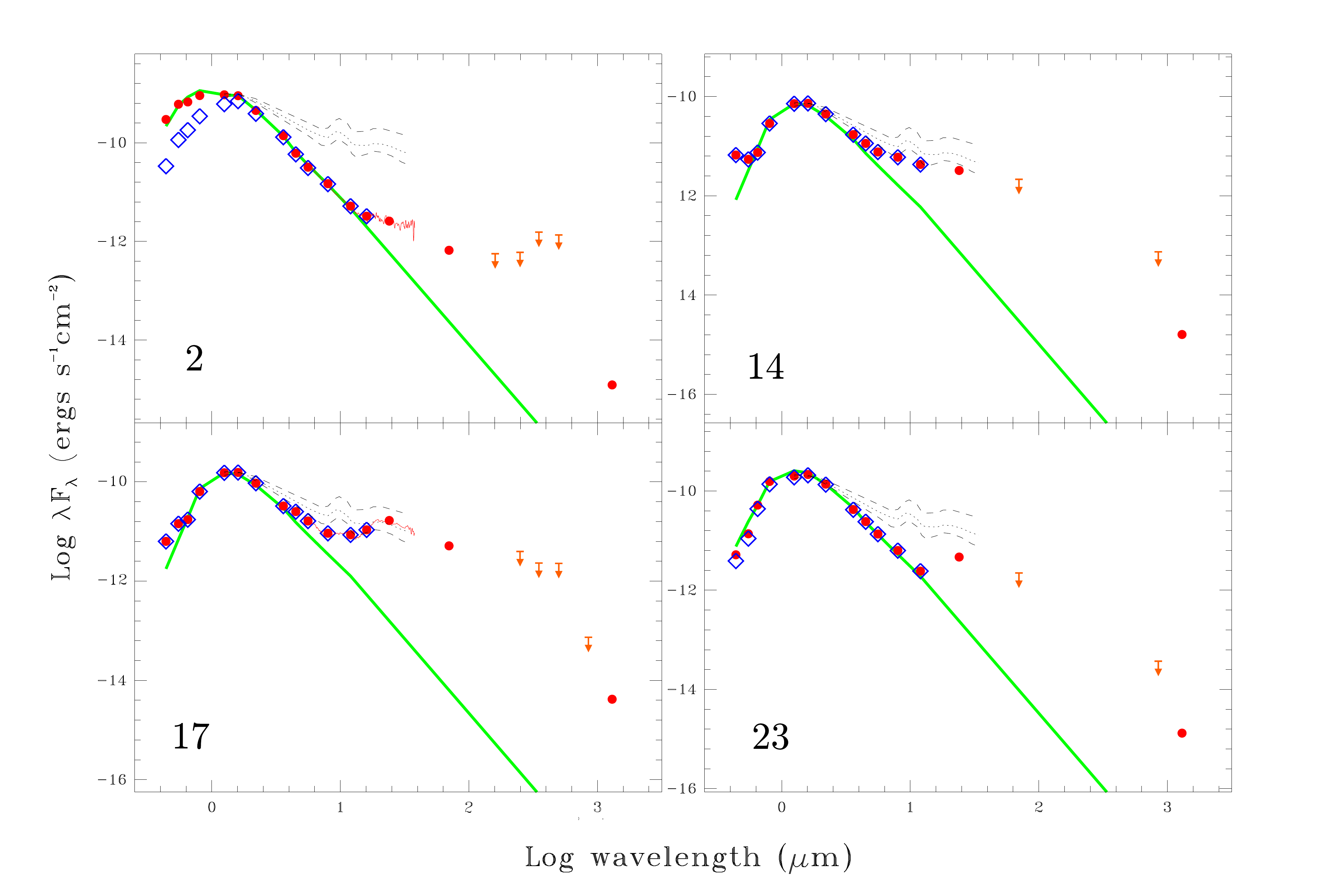}}
        \caption{SEDs of the 4 targets in which 1.3mm flux was detected. The filled circles are archival photometry points, with the upper limits as arrows. The blue diamonds show the flux before being corrected for extinction. The solid green lines represent the stellar photosphere
normalised to the extinction-corrected J band. The dotted lines correspond to the median mid-IR SED of K5–M2 CTTSs calculated by \citet{Furlan2006}, with the dashed
lines as the upper and lower quartiles. The red line corresponds to IRS spectra, where available.
}
        \label{SED_withflux}

\end{figure*}

Only four of our 24 systems have detectable continuum above the 3$\sigma$ level, with an additional tenuous detection in system 8. This system displays weak emission, but which is just below the 3 sigma level and therefore not significant enough to claim a detection.   

\cite{Andrews2005} show that, owing to continuum emission being optically thin at mm wavelengths, the dust mass can be estimated by a simple equation of the form $M_{dust}=C_{\nu} \times F_{\nu}$ where $ C_{\nu}$ is a constant for a given frequency $F_{\nu}$. We adopt the constant derived for 1.3mm by \cite{Cieza2008a}, and use the equation
\begin{equation}   
      M_{\mathrm{dust}}=0.566 \times \left[ \dfrac{F_{\nu}(1300)}{mJy} \left( \dfrac{d}{140 pc} \right)^{2} \right] M_{\oplus}
   \end{equation} 
to estimate the dust mass for our targets, or the dust mass upper limit in the case of a non-detection. This approach has significant uncertainties but is quite standard in the field and therefore allows for meaningful comparisons with previous results. The results of this are displayed in Table \ref{results}. In the case of a non-detection, we quote the 3 sigma value as our upper limit, with this uncertainty being dominated by the uncertainty in the distance to the system. 
 
\subsection{$^{12}$CO(2-1) non-detections}
   
We did not detect $^{12}$CO(2-1) emission for any of the observed
systems suggesting that very little gas remains in the disks.
Determining upper limits on the gas mass that these non-detections
imply is more difficult than for the dust, however,
because the CO emission is generally optically thick and the
conversion to H$_{2}$ (the component that makes up the majority
of the gas mass) is uncertain. This uncertainty is largely due to photo-dissociation and
freeze-out, which depend on the density and temperature
structure of the disk.
To help constrain disk gas masses from CO observations,
\citet{Williams2014} created a grid of models with different density structures that
include a simple prescription for the CO chemistry.
The best constraints on gas masses come from a combination
of the moderately optically thin CO isotopologues
but, because of the sensitivity of these observations,
we can obtain useful limits just by using the $^{12}$CO(2-1) line.
We compare our results to this grid for a stellar mass
of 0.5 \Msun, as this mass is suitable for the stars in our sample {(see Table \ref{results})}, but we leave all other parameters free
(disk mass, {disk radius, radial power-law index}, surface-density power-law index,
temperature profile, and inclination).
For the most massive dust disk in our sample, i.e. System 17,
an ISM gas-to-dust ratio of 100 would imply a gas mass of
$\sim 1\,M_{\rm J}$, which would be readily detectable in
$^{12}$CO(2-1) at our $3\sigma$ detection limit of 90 mJy km\,s$^{-1}$.
We can therefore conclude that the gas-to-dust ratio is
substantially lower than 100 in this disk.
A similar argument applies to system 14.
For the other disks, despite the enhanced effects of photo-dissociation resulting from their lower maximum gas masses of
$M_{\rm gas}\sim 0.1\,M_{\rm J}$, the CO emission would still
be above our detection threshold for 90\% of the models. This suggests
that the gas-to-dust ratio in these disks has also evolved to well below the ISM
value and that the amount of gas in these disks is extremely small.

\subsection{Stellar ages}

To be able to make a comparison with previously studied systems, and to impose limits on disk-clearing timescales, the stellar ages are required. These were obtained by
calculating stellar luminosities and temperatures, and  comparing these to pre-main sequence (PMS) isochrones. Stellar temperatures were estimated by their spectral type, using the scale provided by \citet{Kenyon1995}. The stellar luminosity was calculated by first applying a {de-reddening correction} to each star, as appropriate for its spectral type, again in accordance with the values provided by \citet{Kenyon1995}. The J band magnitude was used as a reference because this band is less affected by extinction than at shorter wavelengths, whilst having little chance of being affected by the flux from any circumstellar material. {The distances in Table \ref{sampleparams} were then used to calculate the stellar radius required to recreate the measured emission, and these radii used to calculate the stellar luminosity. The values of temperature and luminosity were then compared to the PMS isochrones of \citet{Siess2000} to obtain the age.} The results of this process are displayed in Table \ref{results}, {along with the corresponding stellar mass suggested by the PMS isochrones}. Four systems (1, 4, 6, and 7) appeared too under-luminous for their temperature to coincide with any PMS model, and therefore do not have a corresponding value for the age column in Table \ref{sampleparams}.  

Some caution must be taken when interpreting these ages, since the evolutionary tracks for these PMS systems lie close together on the Hertzsprung-Russel diagram. Moreover, the distances to the individual objects have an uncertainty of up to $\sim$15$\%, $ which can introduce a $\sim$30$\%$ uncertainty into the intrinsic luminosity of the object. 
These effects can conspire to result in a high degree of uncertainty in the ages of individual stars. Fortunately, however, the ages of WTTS are considerably easier to determine than those of  CTTS. This is because CTTS are affected by veiling and possess highly heterogeneous photospheres due to their accretion, which makes their intrinsic luminosities and temperature much more challenging to determine \citep{Cieza2005}. In the case of our WTTS, the largest contributor to the uncertainty is probably the distance of the object. 

\subsection{Fractional disk luminosities}

\begin{figure*}
\centering
\includegraphics[width=16.4cm,clip]{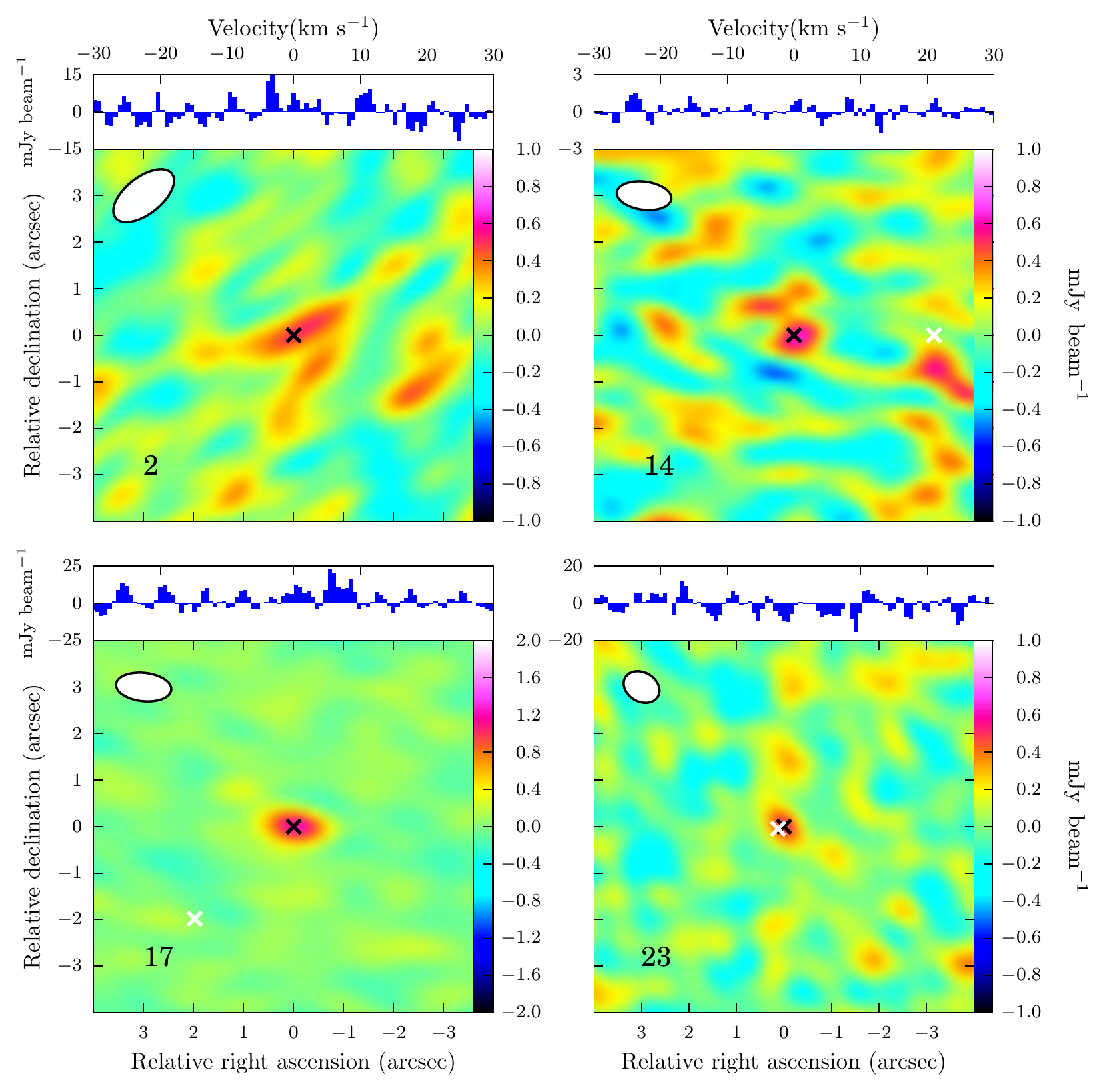}
\caption{Images generated by the CLEAN algorithm for the four systems in which we detected continuum emission. The black cross denotes the position of the target, and the white cross  the approximate position of any potential binary companions. The white ellipse denotes the beam size. The panel above each image displays spectra centred on the $^{12}$CO(2-1) line for the region of continuum emission, which can be seen to contain only noise.}
\label{almadet}
\end{figure*}

A frequently used criterion for distinguishing between protoplanetary and debris disks is the fractional disk luminosity, $L_{D}/L_{\ast}$, which measures the fraction of the stellar radiation that is intercepted and re-emitted by the disk. Typical values for protoplanetary disks are in the region $L_{D}/L_{\ast}\sim 0.1$ \citep[e.g.][]{Cieza2010}, whereas the value for debris disks is much lower, with typical values  $L_{D}/L_{\ast}\lesssim 10^{-3}$ \citep[e.g.][]{Decin2003}. {To determine this quantity for our sample, we attribute all emission from the star-subtracted SED to the disk ($L_{D}$) and  fit this emission to either one or two black bodies, as required. The black body emission at wavelengths longer than $\lambda_{0}$ was modified by a factor of $(   \lambda_{0}/{\lambda}) ^{\beta}$, since, at these wavelengths, the emission from an optically thin debris disk is observed to deviate from a simple black body \citep{Hildebrand1983}. Although $\lambda_{0}$ is often set as a free parameter, the models here lack sufficient photometry to suitably constrain this value. It was therefore assigned a fixed value of 70$\mu$m, which is a reasonable value for such disks \citep{Booth2013}. A typical value of $\beta$, also known as the spectral index, is two for ISM-like material. An equation of the form  
   \begin{equation}   
      S_{v} = \Omega N  \kappa_{0} \left(   \frac{\lambda_{0}}{\lambda} \right) ^{\beta} B_{\nu}(T)
   \end{equation}
was therefore used, where $\Omega$ is the solid angle of the emitting region, N  the column density of dust, $\kappa_{0}$ the opacity of the dust, and $B_{\lambda}(T)$ is the emission of a black body at a temperature T. With values of N, T, $\kappa_{0}$ and spectral index $\beta$ all being allowed to vary, the results become degenerate and therefore cannot be used to infer any specific disk properties}. The total disk flux density however, is well approximated by this simple prescription. The stellar flux density was likewise fit assuming a pure black body spectrum of the star's temperature and normalised to the J band flux. Both flux densities were then integrated according to Simpson's rule and divided to obtain the fractional disk luminosity.  

When disk excess is only detected at one wavelength, the value of $L_{D}/L_{\ast}$ is extremely unconstrained. However, an upper limit was calculated by assuming a value of $\beta$=2 and fitting the black body  to the upper limits from ALMA and/or Herschel. 

\subsection{SEDs and individual system parameters}

\begin{figure*}
    \resizebox{\hsize}{!}{\includegraphics{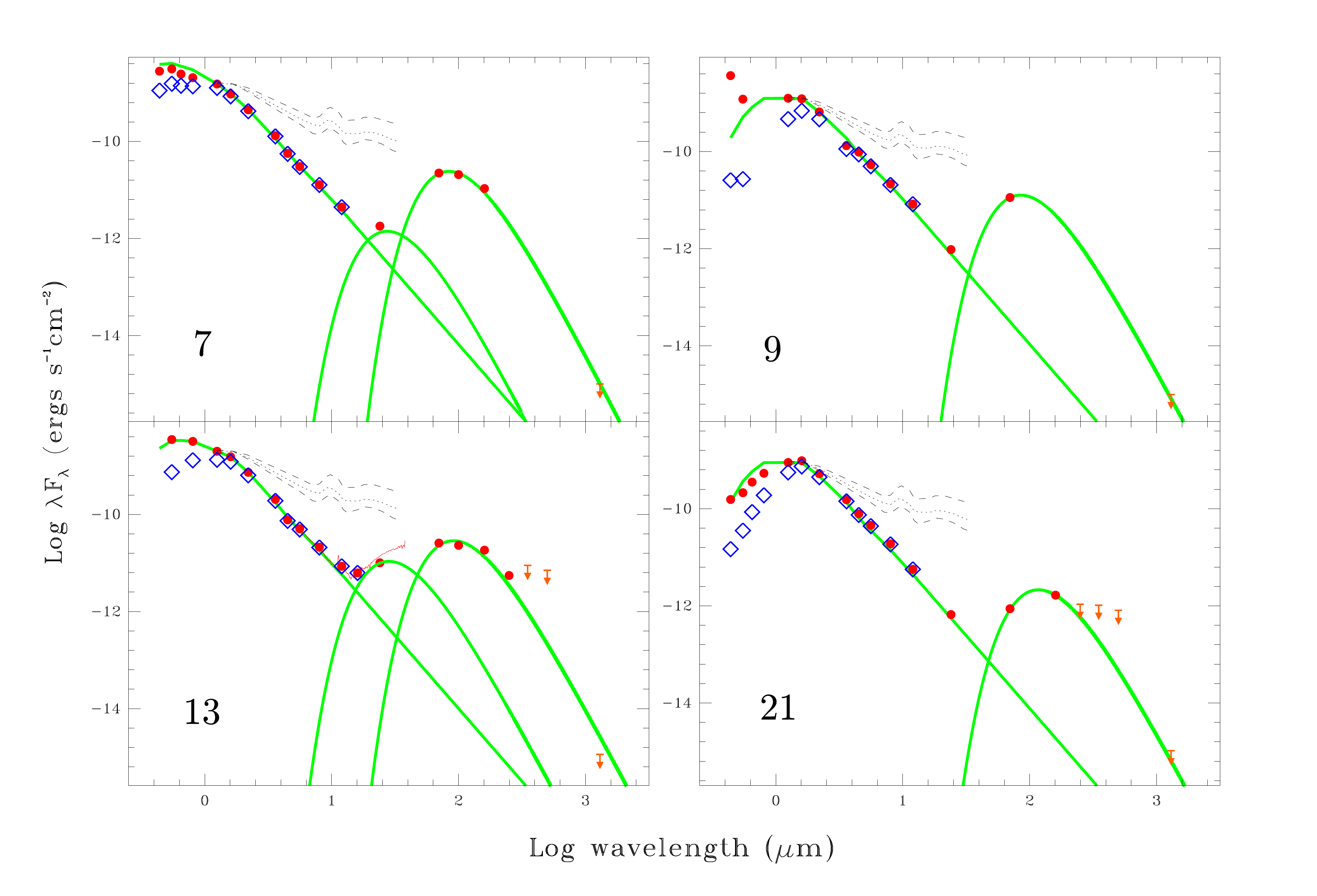}}
        \caption{SEDs for the four systems in our sample that display considerable mid-IR excess, but for which there was no 1.3mm detection. Symbols are identical to Fig. \ref{SED_withflux}, with the addition of modified black body fits to determine if the emission is still consistent with one or two black bodies, taking the ALMA upper limits into consideration.}
        \label{SED_nofluxbutinteresting}

\end{figure*}

\subsection*{Systems with detections (2, 14, 17, 23)}

The SEDs for the four systems in which 1.3mm emission was detected are shown in Fig. \ref{SED_withflux}. System 2 has a fractional disk luminosity that is very low, so is probably a massive debris disk. For the other three systems, however, the mid- to far-IR slope is suggestive of extended dust disks and not the thin belts often seen in debris disks. These three systems are therefore the most likely to be undergoing photoevaporation, and further observations with a deeper detection threshold for the detection of gas will be invaluable. 

An interesting feature of these systems is that they are all believed to be in wide binaries, although there is only single-epoch data available for these systems, so their binary nature has not been confirmed via proper motion \citep{Koehler2008,Merin2008,Nguyen2012}. Assuming that they are indeed binaries, one has to be careful in interpreting where the excess emission originates. {In the case of System 2, the separation is very large ($\geq$1400\,au), and all detected emission can be confidently attributed to the one star}. For System 17, the projected separation is large enough that the photometry short of 10 $\mu$m is resolved (the Spitzer resolution at 8 $\mu$m is $\sim$2''), as is the mm photometry reported in this paper. The 24 and 70 $\mu$m points should be treated with some caution, but since no 1.3 mm emission was detected for this alleged binary companion (see Fig. \ref{almadet}), it is likely that emission at all wavelengths originates from a single circumstellar disk. A similar argument applies to System 14, although the ALMA image does suggest some kind of emission from the approximate location of the binary companion, which may contribute to the 24 $\mu$m flux found for this target as well. Given the wide separations of these targets, with projected separations of $\sim$420 and 560 au, it is unlikely that the tentative binary companion would influence the circumstellar disk evolution considerably. This is supported by \citet{Harris2012}, who found no difference in disk luminosity for disks in wide binaries (separations $\geq$300 au), compared to those around single systems. 

For System 23, the projected separation is too small for any of the detected excess emission to be resolved, and it could therefore originate in either circumstellar disks or a single circumbinary disk. In either case, if the true separation of this system is $\leq$40 au, it is likely that any disks will undergo a very different evolution compared to disks around single stars. For example, observations suggest that binaries of separations $\leq$40 au inhibit the formation of protoplanetary disks \citep{Kraus2012}. However, disk-possessing close binaries can be found, and there is evidence to suggest that the tidal torque from such a binary could even slow down disk evolution \citep{Alexander2012}. This would cause prolonged lifetimes for the disk material, which may  explain this particular detection. However, the binary nature for all systems needs to be confirmed and the sample size increased before conclusions can be drawn about how binarity affects this final stage of a disk's lifetime.  

Systems 14 and 17 are the only systems in our sample that contain excess emission at 12 $\mu$m, suggesting the presence of dust located at small radii. Indeed, the other mid-IR detections of these two systems only deviate slightly from the average disk emission in Taurus. This, coupled with the fact that these two systems have the highest detected dust mass, makes them the most likely to have recently ceased accreting and be in the final stages of gas clearing. As mentioned in Section 3.2, the lack of a CO detection in these systems suggests that if this is the case, the gas-to-dust ratio must be severely depleted. 

\subsection*{{Systems with large mid-IR excess (7, 9, 13, 21)}}

\begin{figure*}
    \resizebox{\hsize}{!}{\includegraphics{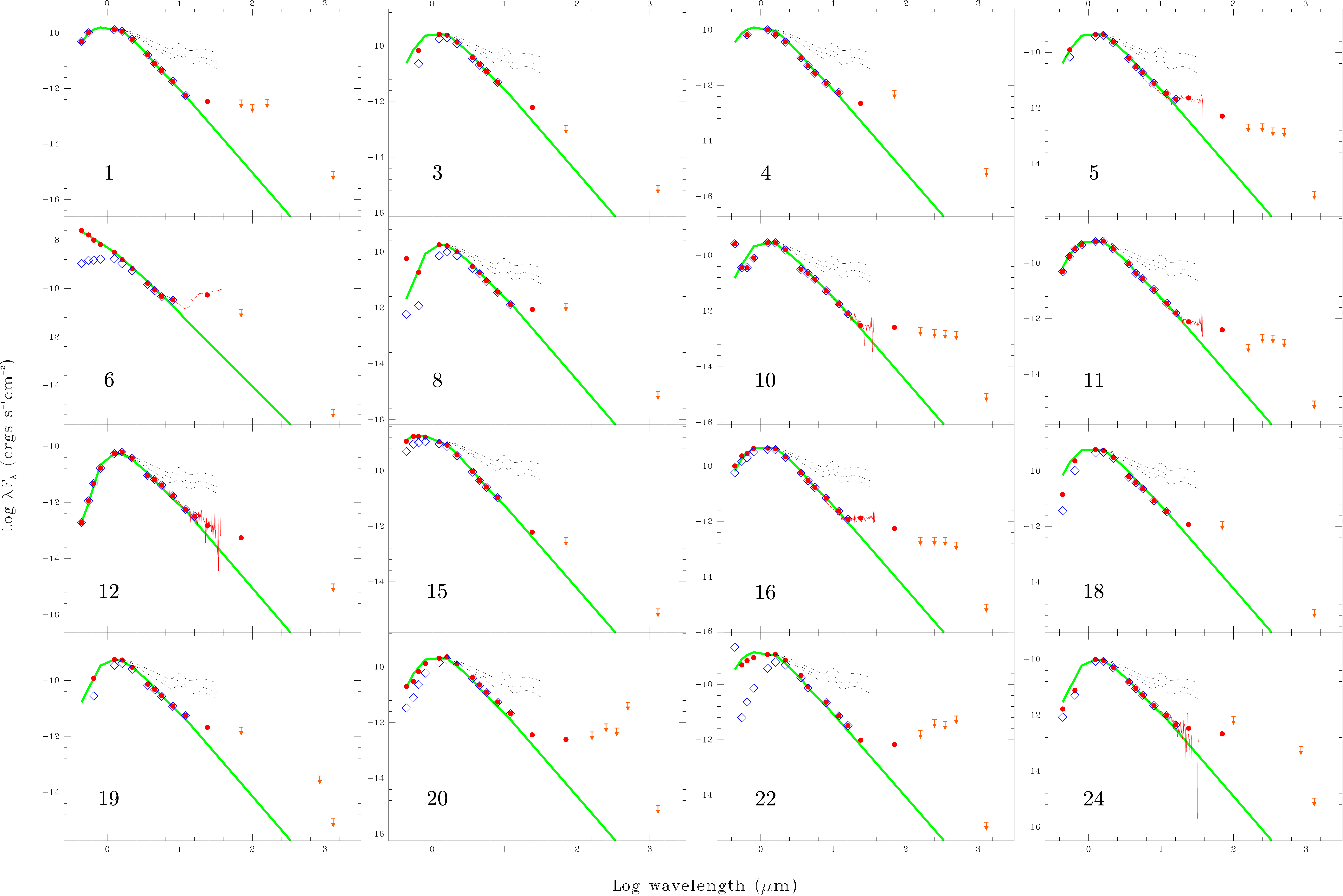}}
        \caption{SEDs for the systems in our sample in which there was no 1.3mm detection, and for which the emission is consistent with a thin belt of material at constant temperature. Symbols are the same as in Fig. \ref{SED_withflux}}
        \label{noflux}

\end{figure*}

Systems, 7, 9, 13, and 21 all have published 70 $\mu$m detection and steep rises in the mid-IR, which  suggests a large population of cold dust (see Fig. \ref{SED_nofluxbutinteresting}). As a result, the ALMA non-detections are rather surprising. We therefore include the black-body fits used to determine their fractional disk luminosity into Fig. \ref{SED_nofluxbutinteresting}, to investigate whether the value for the spectral index must be $\beta\gg$2. All fall below or on the ALMA upper limit with a spectral index $\beta$=2, with the exception of system 13, for which a value of $\beta \geq$ 3.2 was required to satisfy the upper limits. This rather high value of beta suggests that the far-IR points of this system are not associated with the source, and \cite{Matra2012} show that show that emission at wavelengths longer than 70 $\mu$m could be dominated by a source south-west of the system. Accounting for this, we only find  a single black body of T=85K, and $\beta$=2 is required to fit the emission of this system.  

For System 7, our age analysis found no PMS tracks that were consistent with its luminosity, suggesting that it is perhaps not a member of the young cloud and is already on its main-sequence. \cite{Massarotti2005} support this hypothesis by showing that the proper motion of this system is indeed too high to be part of Taurus. The high proper motion makes it likely that this system is, in fact, a foreground main sequence star of a later spectral type with less extinction. 

Systems 9 and 21 both have excesses that begin at 70 $\mu$m, suggesting a population of dust far from the star. System 9 can easily be explained by material in a thin belt and is therefore probably a cold debris disk, albeit a very young one. The 160 $\mu$m point obtained for system 21 requires an even cooler debris disk of $\sim$20K to explain all flux, corresponding to a distance of $\sim$600 au. This is extremely far from the star, and so this is unlikely to  occur. Instead, it is more likely that the 160 $\mu$m point includes contamination from extended emission in the Herschel image from which it was derived \citep{Cieza2012}. In any case, both systems fit with a spectral index of $\beta$=2, which is consistent with dust similar to the ISM.

If the long wavelength emission in Systems 13 and 21 is indeed due to another source, then it is likely that all four of these systems are in their debris phase. This is also apparent from their fractional disk luminosities, which are approximately $10^{-3}$. A deeper search with ALMA would clarify the situation for these uncertain objects since it could both detect and resolve the continuum emission to a much smaller area than is possible with Herschel. System 7 is perhaps the least certain, however, with a relatively high fractional disk luminosity and some uncertainty on its spectral type. If it transpires that it is a late-type star, then the fractional disk luminosity will be increased, and this object would perhaps need to be re-classified. 

\begin{figure*}
    \resizebox{\hsize}{!}{\includegraphics{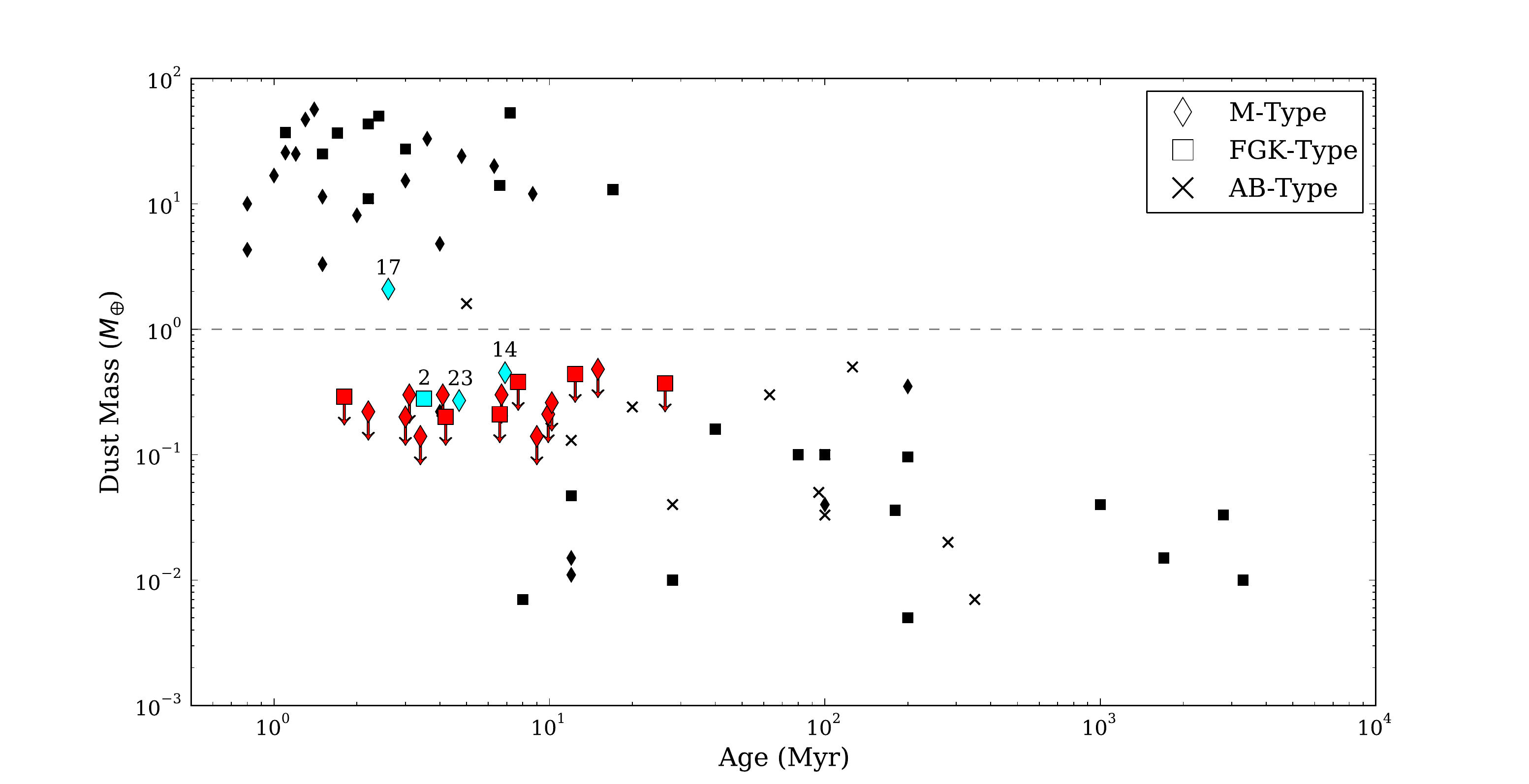}}
        \caption{{Comparison of the dust mass in our systems compared to those found in the literature. Cyan symbols denote the systems in our sample with detected 1.3\,mm flux (numbered for clarity) and red symbols denote those those with upper-limits}. Black symbols that lie above the dotted line correspond to known protoplanetary disks \citep[data taken from][]{Wyatt2003, Ricci2010, Romero2012}. Black symbols in the lower portion correspond to known debris disks \citep[data taken from][]{Greaves2004,Liu2004,Sheret2004,Najita2005,Lestrade2006,Williams2006,Matthews2007}. The horizontal dashed line is used to highlight the difference in dust mass between the populations}
        \label{ages}

\end{figure*}

\subsection*{{Likely debris disks}}
For the remaining systems (Fig. \ref{noflux}), only one or two excess points exist, and the upper limit on the 1.3mm flux is not so restrictive. As a result it is easy to explain the excess as originating from a single temperature black body. The value of $L_{D}/L_{\ast}$ is confined to the debris regime, but it is difficult to class these objects as debris disks without knowledge of their gas content. The argument presented in section 3.2 suggests that, for the majority of disks, CO should have been detected despite the effects of photo-dissociation and our CO non-detections therefore suggest these disks are indeed gas poor. In support of this, previous studies have searched directly for H$_{2}$ tracers in similarly young, low mass systems and found nothing within a few au of the star \citep{Pascucci2006, Ingleby2009}. For example, \citet{Pascucci2006} were able to rule out gas masses above 0.04 \Mjup\, within 3 au of the inner disk radius for the disks in their sample, which includes disks in a similar 5-15 Myr range. They conclude that gas dissipation is very efficient, so it would be surprising if the gas has survived in the low mass systems in our sample. We therefore consider it most likely that the majority of these systems are in their debris stage, containing only thin belts of dust and no gas. Systems 1, 4, and 6, which all appeared too under-luminous to fit a PMS track, are probably also debris disks around background main-sequence stars. There is evidence to suggest that 70 $\mu$m observations are more sensitive to debris disks around such high-temperature objects (Cieza et al. 2008a), allowing them to be detected at larger distances. 

\section{Discussion}
\subsection{WTTS as young debris disks}
{Figure \ref{ages} compares the ages and dust masses of the systems in our sample to those in the literature. Literature values for dust mass were all calculated from optically thin emission in a similar manner to this paper, and ages were estimated using either evolutionary track models \citep{Greaves2004, Sheret2004, Matthews2007, Ricci2010, Romero2012}, Stromgren photometry \citep{Wyatt2003}, lithium lines \citep{Liu2004, Sheret2004, Najita2005, Lestrade2006}, or cluster membership \citep{Williams2006}. The upper limits and detections for our sample clearly show that the dust masses for 23 of the 24 objects lie in the debris disk regime. Furthermore, the fractional disk luminosity for the majority of the objects has a value of $L_{D}/L_{\ast} \leq 3\times10^{-3}$, suggesting they are most probably young debris disks (see Table \ref{results}). Even System 2, in which there is a detection at 1.3 mm, has a value that places it in the debris regime, making it potentially a high-mass debris disk rather than a photoevaporating disk. Systems 14, 17, and 23, on the other hand, have relatively high fractional disk luminosities, which, combined with their 1.3\, mm detections, suggests they are not traditional debris disks and may instead be photoevaporating disks. The lack of gas detection in these systems may also be explained by photoevaporation, although another possibility worth bearing in mind is that they already possessed a reduced gas-to-dust ratio before reaching the WTTS state. Evidence for this possibility has recently been found by \citet{Ansdell2015}, in which a reduced gas-to-dust ratio of $\lesssim$2 was found in an accreting system, potentially as a result of dust filtration \citep{Rice2006}.}

Taking the broad definition of a debris disk, where they are defined as gas-poor, geometrically thin dust disks at a uniform temperature, then the previously youngest known debris disks have ages in the range of { 3-5 Myr \citep{Pascucci2006,Chen2014,Rigliaco2015}. The systems observed here, however, generally have younger ages, with 10 systems lying within the age range $1.8_{-0.5}^{+1.2}$ to $4.7_{-0.7}^{+1.0}$ Myr}. Even accounting for the intrinsic uncertainty in these ages, they are significantly lower than previously detected debris disks, and this raises questions as to why such young disks have not been observed before. This is quite possibly the result of observational bias, since debris-disk emission is intrinsically faint and, as such, observations are limited to nearby stars. The arrival of ALMA, however, now allows us to open this parameter space to  young clusters at much greater distances of $\geq$ 100 pc and our results clearly hint that rapid evolution into the debris phase is possible even for late spectral types. 
If confirmed, then these systems will lower the minimum age of debris disks and indicate a wider range of time scales for protoplanetary disk evolution.  

A more restrictive definition of debris disks describes them as containing second-generation dust, which is formed through a continuous process of collisions between planetesimals, and subsequent removal through radiation pressure and Poynting-Robertson drag \citep{Wyatt2008}. Using this definition, it is unclear whether the disks in our sample can be classed as debris. \citet{Wahhaj2010} suggest that dust around WTTS is primordial, based on the lower fraction of IR-WTTS found when comparing off-cloud to  on-cloud sources. They suggest that this is caused by dissipating primordial dust since there is a link between separation from the cloud and age, with the older WTTS located at increased separation from their parental cloud. This interpretation is still rather uncertain however, because both on- and off-cloud sources in the survey have a wide variety of ages with considerable overlap, resulting in only a weak trend in age as a function of separation. 
Instead, the majority of both observational evidence and photoevaporation models indicate that dust in the inner and outer regions of disks dissipates more or less simultaneously \citep[e.g.][]{Andrews2005,Alexander2007}, and most models of photoevaporation do not allow a considerable amount of dust to be left behind. This therefore implies that the debris disks seen around WTTS contain second generation dust. It has also been suggested that a large portion of the dust in protoplanetary disks could even be second generation, based on the observation that the growth of grains from micron to metre sizes can occur very rapidly \citep{Dominik2007}. This would quickly reduce the dust mass of protoplanetary disks inferred from IR observations, and yet this value remains fairly constant for a range of disk ages \citep{Natta2007}. A mechanism of dust replenishment is therefore required in protoplanetary disks \citep{Dullemond2005}, and we consider it likely that the debris disks in our sample contain second-generation dust as well. 

The observed fraction of the WTTS population that display an excess ($\sim20\%$) is similar to the fraction of debris disks found around young FGK type stars \citep[between 10-16\%;][]{Hillenbrand2008, Trilling2008}. Although it is often assumed that IR-WTTS evolve into those without an excess, an alternative possibility is that these IR-WTTS mostly contain debris disks of second-generation origin, and these debris disks can then persist into their main sequence lifetime. The WTTS without excess would then make up an entirely different population, and the difference between these two populations could be that the former has a method of stirring the disk, which is a requirement for the collisional cascades that form debris disks. Stirring via stellar flybys \citep{Kenyon2002} and self-stirring via 1000 km-sized planetesimals \citep{Kenyon2010} are both possible causes for this. Perhaps the most exciting explanation for WTTS with debris disks, however, is that these systems have formed giant planets capable of stirring the disk \citep{Mustill2009}. This would require giant planets orbiting at a few au, and the occurrence for such bodies around FGK stars has been estimated from radial velocity surveys at between 12\% to 22\% \citep{Lineweaver2003,Marcy2005,Cumming2008}. The similarity between this occurrence rate and the number of WTTS systems that possess debris disks makes such systems  excellent places to perform planet searches. 

\subsection{Implications for photoevaporation models} 
Some models of X-ray photoevaporation predict a sample of 'relic' transition disks with large cavities and high dust masses that persist for $\geq$10 Myr \citep{Owen2011}. Our findings add to the growing body of research going against this prediction \citep[e.g.][]{Cieza2010, Cieza2012, Mathews2012}, since the vast majority of our IR-WTTS have no detectable 1.3 mm emission despite their ages being under 10 Myr. Some of the more recent X-ray photoevaporation theories introduce a mechanism they call "thermal sweeping", which disperses the remaining material in these massive dust disks in a small fraction (1-3\%) of the disk's total lifetime \citep{Owen2013} and removes the prediction of these relic disks. Likewise, the EUV models of photoevaporation predict short time scales for final disk clearing. EUV models predict much lower photoevaporation rates in general, but are also only capable of forming transition disks when viscous accretion has already cleared a large amount of material. The resulting disks predicted by EUV models therefore have lower mass than in the X-ray case, and clearing can progress on time scales of between 1\% to 10\% of the disk's lifetime \citep{Alexander2007}.  

Previous surveys of WTTS have found that $\sim20\%$ of WTTS display an excess that, when compared to the number of CTTS in the same regions, suggests that the disks around WTTS persist for 10\% to 20\% of the disk lifetime before moving into a diskless state \citep{Cieza2007, Wahhaj2010}. This percentage is somewhat higher than  predicted by photoevaporation models, but its derivation assumes that all IR-WTTS follow the same evolution, moving from a CTTS, to an IR-WTTS, and finally to a diskless state. If, as outlined above, the WTTS with debris disks are not in transition, then the apparent rarity of photoevaporating disks will significantly lower this percentage and may bring it more in line with photoevaporation models. The small number of detections and lack of gas confirmation in our potential photoevaporating disks does not allow accurate estimation of this corrected percentage, but a wider survey of these photoevaporating disks may confirm this tentative result.

\subsection{Comparison to known debris disks with gas}
In recent years, a small population of debris disks have been found with detectable gas, leading to some debate as to its origin. As with the dust, the origin of this gas is believed to be either primordial or formed through collisions of icy comet-like objects. One such system is 49 Ceti, with spectral type A1, a dust mass of $\sim$0.3 M$_{\oplus}$ and a $^{12}$CO(2-1) integrated intensity of 2.0 Jy km\,s$^{-1}$ \citep{Hughes2008}. Its age has proved difficult to determine since it is not obviously a member of any associations, but \citet{Thi2001} believe it is a PMS star with an age of 8 Myr, opening up the possibility that it is a high-mass analogy to the stars in our sample. If 49 Ceti were at the distance of Taurus, however, we would expect to detect this level of CO. Likewise, the 30 Myr-old, A4 type system HD\,21997 is classed as a debris disk, containing only 0.09 M$_{\oplus}$ of dust, and yet it displays CO emission that we would have conclusively detected in our survey \citep{Kospal2013}. If these systems are truly harbouring primordial gas, then the evolution for A-type stars must be drastically different to late-type stars to allow them to retain such a large quantity of gas at such low dust masses. Alternatively, the gas is secondary and both A type stars and most of the late-type stars in this survey lose their primordial gas by ages of $\lesssim$10Myr.

\section{Conclusion}

All the above sources in our sample are beyond the stage of active gas accretion. Their SEDs are suggestive of a depleted dust mass, and here we confirm this, either with their low 1.3 mm flux or with their non-detection from ALMA. The dust for all systems must, therefore, have either been largely removed or agglomerated into larger particles. The non-detection of CO lines in all systems is suggestive of a similar fate for the gas, which was probably removed by photoevaporation. Although photo-dissociation models for the disks studied here suggest that the CO abundance will be lowered slightly, for the majority of disks CO should have remained detectable. {It is therefore probable that the gas-to-dust ratio has evolved in the majority of these disks to a value well below that of the ISM. The non-detection of CO is particularly surprising for Systems 14 and 17, whose SEDs strongly suggest large dust reservoirs. The depleted gas-to-dust ratio in these two systems may therefore be the result of a late stage of photoevaporation, or else they already possessed a reduced gas-to-dust ratio before reaching the WTTS state.}

For those systems in which there was no 1.3 mm detection, it is probable that they are free from gas and contain dust masses and distributions similar to debris disks.  
This is apparent from a comparison of their fractional disk luminosities and dust mass upper-limits to that of known debris disks, as both lie in the debris disk regime. These systems, however, are clearly much younger than the majority of debris disks allowing for more strict constraints on debris disk formation time scales than ever before. A deeper study with ALMA will be invaluable to determine conclusively their evolutionary state, as well as to confirm the dust masses of the suspected debris disks.    

\begin{acknowledgements}

{The authors would like to thank the anonymous referee for the useful comments.} AH, MRS, CC, HC, and LC acknowledge support from the Millennium Nucleus RC130007 (Chilean Ministry of Economy). MRS, CC and LC also acknowledge support from FONDECYT (grants 1141269, 3140592, 1440109), and HC acknowledges support from ALMA/CONICYT (grants 31100025 and 31130027). RDA acknowledges support from The Leverhulme Trust though a Philip Leverhulme Prize, and from the Science \& Technology Facilities Council (STFC) through Consolidated Grant ST/K001000/1. JPW is supported by the  NSF, through grant AST-1208911 and NASA, through grant NNX15AC92G. In addition, the authors would like to thank the organisers of MAD workshop in Santiago, which made this collaborative science result possible. This paper makes use of the following ALMA data: ADS/JAO.ALMA\#2012.1.00350.S, ADS/JAO.ALMA\#2011.0.00733.S. ALMA is a partnership of ESO (representing its member states), NSF (USA), and NINS (Japan), together with NRC (Canada) and NSC and ASIAA (Taiwan) and KASI (Republic of Korea), in cooperation with the Republic of Chile. The Joint ALMA Observatory is operated by ESO, AUI/NRAO, and NAOJ.

\end{acknowledgements}

\bibliographystyle{aa}
\bibliography{Disks}




\end{document}